\documentclass[11pt,reqno]{amsart}
\usepackage{amsfonts,amsmath,amssymb}

\newtheorem{thm}{Theorem}[section]

\usepackage{graphicx}
\usepackage{epstopdf}

\title{Quantum Calogero--Moser systems: a view from infinity}

\author{A.N. Sergeev}\address{Department of Mathematical Sciences,
Loughborough University, Loughborough LE11 3TU, UK}
\email{A.N.Sergeev@lboro.ac.uk}

\author{A.P. Veselov}
\address{Department of Mathematical Sciences,
Loughborough University, Loughborough LE11 3TU, UK and Department of Mathematics and Mechanics, Moscow State University, Moscow, 119899, Russia}
\email{A.P.Veselov@lboro.ac.uk}

\begin{document}

\begin{abstract}
Various infinite-dimensional versions of Calogero--Moser operator are discussed in relation with the theory of symmetric functions and representation theory of basic classical Lie superlagebras. This is a version of invited talk given by the second author at XVI International Congress on Mathematical Physics in Prague, August 2009.
\end{abstract}
\maketitle

\section{Introduction}

Calogero--Moser systems play truly exceptional role in the modern theory of integrable systems
(see e.g. \cite{CMS} for a variety of both mathematical and theoretical physics problems they are related to). They have natural physical interpretation, describing the interaction of $N$ particles with equal masses on the line with the inverse square potential or, in Sutherland's version, with the inverse $\sin^2$ potential. 

They admit natural generalizations related to root systems and simple Lie algebras \cite{OP}, and, at the quantum level only, also non-symmetric integrable versions called {\it deformed Calogero-Moser systems} \cite{CFV}, which turned out to be related to basic classical Lie superalgebras \cite{SV}.
In particular, in the case of Lie superalgebra $\mathfrak{sl} (m,n)$ we have two groups of particles with two different masses with the parameters of interaction inside the groups and between them being "tuned" in a very special way (see \cite{SV}).

It turned out that these mysterious deformations can be clearly "seen from infinity". For this one should first explain what is the analogue of the Calogero--Moser operators when $N=\infty.$ A proper framework is given by the theory of symmetric functions and goes back to Stanley \cite{Stanley} and Macdonald \cite{Ma}, who were inspired by the work of H. Jack. It is interesting that Jack did his work on what is now called {\it Jack polynomials} around 1970 - almost at the same time as the pioneering work by Calogero and Sutherland, but a close relation between these two important developments was not recognized until much later. 

The infinite-dimensional point of view also leads naturally to the notion of {\it super Jacobi polynomials}. Their specialized versions turned out to coincide with suitable Euler supercharacters of the orthosymplectic Lie superalgebras \cite{Ser, SV4}, which is one more evidence of the deep link between Calogero--Moser systems and representation theory.

\section{Calogero--Moser operators in infinite dimension}

Consider the following Calogero--Moser--Sutherland operator ({\it CMS operator}):
$$L_k^{(N)} = \sum_{i=1}^N
\frac{\partial^2}{\partial
x_{i}^2}-\sum_{i<j}^N \frac{2k(k+1)}{\sinh
^2(x_{i}-x_{j})},$$ 
where $k$ is a parameter and for convenience we have changed the sign of the operator. 
Its gauged version $ {\mathcal L}_{k}^{(N)}=\Psi_0^{-1} (L_N-\lambda_0) \Psi_0$ with $\Psi_0= \prod_{i<j}^N \sinh^{-k}(x_i-x_j),\,\,$  
$\lambda_0= k^2N(N-1)/4$ in the exponential coordinates 
$z_i = e^{2x_i}$ has the form
\begin{equation}
\label{CM}
 {\mathcal L}_{k}^{(N)}=\sum_{i=1}^N
\left(z_{i}\frac{\partial}{\partial
z_{i}}\right)^2-k\sum_{ i < j}^N
\frac{z_{i}+z_{j}}{z_{i}-z_{j}}\left(
z_{i}\frac{\partial}{\partial z_{i}}-
z_{j}\frac{\partial}{\partial
z_{j}}\right).
\end{equation}
It preserves the algebra of symmetric polynomials $\Lambda_{N} = \mathbb C[z_{1},\dots, z_{N}]^{S_N}.$ This algebra has a natural infinite-dimensional version $\Lambda=\varprojlim\Lambda_{N}$
defined as the inverse limit in the category of graded algebras \cite{Ma}; the elements of $\Lambda$ are called {\it symmetric functions}. 

The power sums
$p_l = z_1^l + z_2^l + \dots,\,\, l =1,2,\dots$ give a convenient set of free generators of this algebra, which can be considered also as the coordinates in the corresponding infinite-dimensional space. 
We have a natural homorphism $\varphi_N: \Lambda \rightarrow \Lambda_N,$ sending $p_l$ to its $N$-dimensional analogue $p_l=z_1^l+\dots +z_N^l.$
There is a problem with
$p_0=1+1+\dots$, which in finite-dimensional situation is just the dimension $N$, but in infinite dimension does not make sense. The solution is to consider $p_0$ as a formal {\it additional parameter}, which our operators may depend on. Under the homomorphism $\varphi_N$ it must be specialized to dimension $N.$ 

An infinite-dimensional analogue of CMS operator $\mathcal L_{k,p_0}^{(\infty)}: \Lambda \to \Lambda$ is defined as the unique second order differential operator polynomially dependent on $p_0$, such that for all $N=1,2,\dots$ and $p_0=N$   the following diagram is commutative
$$
\begin{array}{ccc}
\Lambda&\stackrel{{\mathcal
L}_{k,p_0}^{(\infty)}}{\longrightarrow}&\Lambda\\ \downarrow
\lefteqn{\varphi_{N}}& &\downarrow \lefteqn{\varphi_{N}}\\
\Lambda_{N}&\stackrel{{\mathcal
L}_k^{(N)}}{\longrightarrow}&\Lambda_{N} \\
\end{array}.
$$
The operator $\mathcal L_{k,p_0}^{(\infty)}$ has the following explicit form in power sums (see \cite{Stanley, Awata, SV5}):
  \begin{equation}\label{inf1}
{ \mathcal L}_{k,p_0}^{(\infty)}=\sum_{a,b>0}p_{a+b}\partial_{a}\partial_{b}-k\sum_{a,b>0}p_{a}p_b \partial_{a+b}- kp_0 \sum_{a>0} p_{a} \partial_{a} +(1+k)\sum_{a>0}a p_a\partial_a,
\end{equation}
where $\partial_a = a\frac{\partial}{\partial p_a}.$

This form reveals an important {\it duality}
 \begin{equation}\label{sym}
\theta^{-1} \circ \mathcal L_{k,p_0}^{(\infty)} \circ \theta = k \mathcal L_{k^{-1}, k^{-1}p_0}^{(\infty)},
 \end{equation}
where $$
\theta: p_a \rightarrow k  p_a, k  \rightarrow k^{-1}.$$
There is also a remarkable symmetry between the first and the second terms, so that if  following \cite{Awata} we define $\tilde \partial_a=-\frac{a}{k} \frac{\partial}{\partial p_a}$ the operator is invariant under swapping $p_a$ and $\tilde \partial_a$ ({\it Fourier duality}).
 
Note that $\theta$ changes the parameter $p_0$, which means that it does not work in the finite dimensions. This fact was known already to Stanley and Macdonald, who probably were the first to discover this duality (see \cite{Stanley, Ma}).
They have actually used a {\it stable version} of the CMS operator by subtracting the momentum operator $P= \sum z_i \frac{\partial}{\partial z_i} = \sum p_a \partial_a$ with dimension dependent coefficient $(N-1)$. These stabilized CMS operators can be lifted to infinite dimension without introducing extra parameter $p_0,$ which corresponds to the fact that $p_0$ appears in (\ref{inf1}) only as a coefficient at $P$, 
which is just another quantum integral of the system.

However, already for the rational Calogero--Moser operator
$$
L_k^{(N)}=\sum_{i=1}^N
\frac{\partial^2}{\partial
x_{i}^2}-\sum_{i<j}^N \frac{2k(k+1)}{(x_{i}-x_{j})^2}
$$
the stabilization trick is not possible: its infinite-dimensional version has the form \cite{SV5}
 \begin{equation}\label{inf1ratio}
{ \mathbf L}_{k,p_0}^{(\infty)}=\sum_{a,b\geq 1}p_{a+b-2}\partial_{a}\partial_{b}-k\sum_{a,b\geq 0}p_{a}p_b \partial_{a+b+2} +(1+k)\sum_{a \geq 2}(a-1) p_{a-2}\partial_a.
\end{equation}
The same is true for the rational $B_N$ Calogero-Moser operator 
$$ L^{N}_{k,l}= \Delta_N -\sum_{i<j}^{N}\left(\frac{2k(k+1)}{(x_{i}-x_{j})^2}+\frac{2k(k+1)}{(x_{i}+x_{j})^2}\right) -\sum_{i=1}^N
\frac{l(l+1)}{x_{i}^2},$$
with 2 parameters $k,l.$ The corresponding infinite-dimensional analogue (after a convenient division by 4) in the coordinates $z_i=x_i^2$ has a very similar form  \cite{SV5}: 
$${ \mathcal B}_{k,l,p_0}^{(\infty)}=\sum_{a,b\geq 1}p_{a+b-1}\partial_{a}\partial_{b}-k\sum_{a,b\geq 1}p_{a}p_b \partial_{a+b+1} +(1+k)\sum_{a \geq 1}a p_{a-1}\partial_a $$
 \begin{equation}\label{inf2}
- (2k p_0+l+1/2)\sum_{a \geq 1}p_{a-1}\partial_a + kp_0^2 \partial_1.
\end{equation}
In the trigonometric $BC_N$ case we have the operator $L^{N}_{k,p,q}=$
$$\Delta_N -\sum_{i<j}^{N}\left(\frac{2k(k+1)}{\sinh^2(x_{i}-x_{j})}+\frac{2k(k+1)}{\sinh^2(x_{i}+x_{j})}\right) -\sum_{i=1}^N
\left(\frac{p(p+2q+1)}{\sinh^2x_{i}}+ 
\frac{4q(q+1)}{\sinh^22x_{i}}\right),$$
depending on 3 parameters $k,p,q.$
The $BC_{\infty}$ version of the CMS operator has the form \cite{SV3}:
$$
 \mathcal{L}^{(\infty)}_{k,p,q,h}=\sum_{a,b>0}(p_{a+b}+2p_{a+b-1})\partial_{a}\partial_{b}-k\sum_{a=2}^{\infty}\left[\sum_{b=0}^{a-2}p_{a-b-1}(2p_{b}+p_{b+1})\right]\partial_{a}
$$
\begin{equation}
\label{bcinf}
+\sum_{a=1}^{\infty}\left[(a+k(a+1)+2h)p_{a}+(2a-1+2ka+2h-p)p_{a-1}\right]\partial_{a},
\end{equation}
where the additional parameter $h$ is related to $p_0$ as 
$h=-(kp_0+\frac{1}{2}p+q).$ 
This form is invariant under three involutions (dualities) \cite{SV3}. One of them is an extension of the duality $k \rightarrow k^{-1}$ from the previous case. It acts on the other parameters as
\begin{equation}
\label{bcrel}
2\hat h-1 = k^{-1}(2h-1),\,
\hat p=k^{-1}p, \,\, (2\hat q+1)=k^{-1}(2q+1).
\end{equation}
The same relations first appeared in the formulas for the $BC(m,n)$ deformed CMS operators related to the orthosymplectic Lie superalgebras $\mathfrak{osp}(m, 2n)$ \cite{SV}. We are going to see now that this is not a mere coincidence.

\section{Deformed CMS operators and classical Lie superalgebras seen from infinity}

We restrict ourselves by the $BC$ case (see the $A$ case in \cite{SV1}).
The $BC(m,n)$ deformed CMS operators have the following form \cite{SV}: 
\begin{eqnarray}
\label{bcnm} L^{m,n}& =& \Delta_m
+k \Delta_n -\sum_{i<j}^{m}\left(\frac{2k(k+1)}{\sinh^2(x_{i}-x_{j})}+\frac{2k(k+1)}{\sinh^2(x_{i}+x_{j})}\right)\nonumber \\& &
-\sum_{i<j}^{n}\left(\frac{2(k^{-1}+1)}{\sinh^2(y_{i}-y_{j})}+\frac{2(k^{-1}+1)}{\sinh^2(y_{i}+y_{j})}\right)
\nonumber
\\& & -\sum_{i=1}^{m}\sum_{j=1}^{n}\left(\frac{2(k+1)}{\sin^2(x_{i}-y_{j})}+
\frac{2(k+1)}{\sinh^2(x_{i}+y_{j})}\right) +\sum_{i=1}^m
\frac{p(p+2q+1)}{\sinh^2x_{i}} \nonumber \\& & +\sum_{i=1}^m
\frac{4q(q+1)}{\sinh^22x_{i}} -\sum_{j=1}^n \frac{k
r(r+2s+1)}{\sinh^2y_{j}}-\sum_{j=1}^n \frac{4k
s(s+1)}{\sinh^22y_{j}},
\end{eqnarray}
where the parameters $k,p,q,r,s$ satisfy the
following relations (cf. formula (\ref{bcrel})):
$$
r=k^{-1}p,\quad 2s+1=k^{-1}(2q+1).
$$

Consider the algebra $\Lambda_{m,n,k}$ consisting of polynomials, which are symmetric in $u_{1},\dots,u_{m}$ and $v_{1},\dots, v_{n}$ separately  and satisfy the conditions
$$
u_{i}\frac{\partial f}{\partial u_{i}}-kv_{\alpha}\frac{\partial f}{\partial u_{\alpha}}=0
$$
on the hyperplanes  $u_{i}=v_{\alpha}$ for all $i=1,\dots,m$ and $\alpha=1,\dots, n.$ For generic values of parameter $k$ this algebra is generated by the {\it deformed power sums} 
$$p_a(u,v,k)= \sum_{i=1}^m u_{i}^a+k^{-1}\sum_{\alpha=1}^{n}v_{\alpha}^a,\,\, \:a=1,2,\dots.$$
The claim is that the $BC(m,n)$ deformed CMS operators are the restrictions of $BC_{\infty}$ CMS operator (\ref{bcinf}) onto the corresponding subvariety $\mathcal D(m,n,k) = Spec \, \Lambda_{m,n,k}.$
More precisely, let $\varphi_{m,n}: \Lambda\longrightarrow\Lambda_{m,n,k}$ be the restriction homomorphism
defined by $\varphi_{m,n} (p_{a})= p_a(u,v,k).$

\begin{thm}  [\cite{SV3}]
The following diagram is commutative for $h=-km-n-\frac{1}{2}p-q$ and generic values of parameter $k$:
\begin{equation} \label{commdia}
\begin{array}{ccc}
\Lambda&\stackrel{{\mathcal L^{(k,p,q,h)}}_{}}{\longrightarrow}&\Lambda
\\ \downarrow \lefteqn{\varphi_{m,n}}& &\downarrow \lefteqn{\varphi_{m,n}}\\
\Lambda_{m,n,k}&\stackrel{{\mathcal
L}^{(m,n)}}{\longrightarrow}&\Lambda_{m,n,k} \\
\end{array},
\end{equation}
where ${\mathcal
L}^{(m,n)}$ is a gauged version of the deformed CMS operator (\ref{bcnm}) rewritten in the coordinates 
$u_i=2\sinh^2x_i, \, v_j=2\sinh^2y_j.$
\end{thm}

A remarkable fact is that these are essentially all possible restrictions of $BC_{\infty}$ CMS operator  (see Theorem 4.6 in \cite{SV3}), which shows the exceptional role of the deformed CMS operators.

The eigenfunctions $\mathcal J_{\lambda}(x; k,p,q,h)$ of the $BC_{\infty}$ operator (\ref{bcinf}) are labelled by partitions $\lambda$ and  known as {\it Jacobi symmetric functions}.
Their images $$\mathcal SJ_{\lambda}(u,v; k,p,q)=\varphi_{m,n}(\mathcal J_{\lambda}(x; k,p,q,h)),$$
where $h=-km-n-\frac{1}{2}p - q$, are called {\it super Jacobi polynomials.} 
It turns out that their specialized versions have a natural interpretation in the representation theory as the {\it Euler supercharacters} of  the orthosymplectic Lie superalgebras
$\mathfrak{osp}(m, 2n).$

There is a classical construction due to Borel, Weil and Bott of the irreducible representations of the complex semisimple Lie groups $G$ in terms of the cohomology of the holomorphic line bundles over the corresponding flag  varieties $G/P$. In the Lie supergroup case this leads to a virtual representation given by the Euler characteristic $$\mathcal E_{\lambda} = \sum (-1)^i H^i(G/P, \mathcal O_{\lambda})$$
for certain sheaf cohomology  groups; the corresponding supercharacter $E_{\lambda}$ is called Euler supercharacter (see \cite{Ser}). 

\begin{thm}[\cite{SV4}]
 For special choices of parabolic subgroup $P$ the Euler supercharacters of $\mathfrak{osp}(2m+1, 2n)$
 coincide with specialized super Jacobi polynomials
$$E_{\lambda} =  \mathcal SJ_{\lambda}(u,v; -1, -1, 0).$$
\end{thm}

A similar fact holds for Lie superalgebra $\mathfrak{osp}(2m, 2n)$ and  $\mathcal SJ_{\lambda}(u,v; -1, 0, 0),$ but the analogue of this result for the Lie superalgebra $\mathfrak{sl}(m,n)$ is still to be found (see \cite{SV5} for further discussion).

\end{document}